\documentclass[12pt]{article}
\usepackage{graphicx}
\newcommand{\dbox}{\,\raise2pt\hbox{\fbox{\rule{2.5pt}{0pt}\rule{0pt}{2.5pt}}}\,}
\newcommand{\qed}{\,\raise0pt\hbox{\mbox{\rule{6.5pt}{6.5pt}}}}

\begin{document}
\setlength{\baselineskip}{7mm}

\begin{titlepage}
 \begin{normalsize}
  \begin{flushright}
        UT-Komaba/22-1\\
        July 2022
 \end{flushright}
 \end{normalsize}
 \begin{LARGE}
   \vspace{1cm}
   \begin{center}
     Quantizing multi-pronged open string junction\\
   \end{center}
 \end{LARGE}
  \vspace{5mm}
 \begin{center}
    Masako {\sc Asano}$^{\dagger}$ 
            \hspace{3mm}and\hspace{3mm}
    Mitsuhiro {\sc Kato}$^{\ddagger}$ 
\\
      \vspace{4mm}
        ${}^{\dagger}${\sl Faculty of Science and Technology}\\
        {\sl Seikei University}\\
        {\sl Musashino-shi, Tokyo 180-8633, Japan}\\
      \vspace{4mm}
        ${}^{\ddagger}${\sl Institute of Physics} \\
        {\sl University of Tokyo, Komaba}\\
        {\sl Meguro-ku, Tokyo 153-8902, Japan}\\
      \vspace{1cm}

  ABSTRACT\par
 \end{center}
 \begin{quote}
  \begin{normalsize}

Covariant quantization of multi-pronged open bosonic string junction is studied beyond static analysis. Its excited states are described by a set of ordinary bosons as well as some sets of twisted bosons on the world-sheet. The system is characterized by a certain large algebra of twisted type which includes a single Virasoro algebra as a subalgebra. By properly defining the physical states, one can show that there are no ghosts in the Hilbert space.

\end{normalsize}
 \end{quote}

\end{titlepage}
\vfil\eject

\section{Introduction}
Since the 1990's, when D-branes and various string dualities were found, string junctions have been studied by many authors in the context of superstrings and M-theory.\footnote{We only cite lectures~\cite{Schwarz:1996bh,Johnson:2000ch} that explain their essential points since there are so many papers and we cannot list them all.} These analyses mainly focused on the static properties such as BPS conditions or stability, with a few exceptions (e.g.~\cite{Rey:1997sp,Callan:1998sf}.) String junctions are dynamical objects formed by dynamical strings, so that one can naturally ask their dynamical properties such as spectrum of their excited states and other quantum features beyond the static properties.

Going back to the 1970's, some earlier works which studied classical motions~\cite{Artru:1974zn,Collins:1975yt,Kiriyama:1976ue} and simple-minded quantization~\cite{Sundermeyer:1976df} of string junctions appeared. In those days string junction was considered as a model of baryon. They naively tried to quantize 3-string junction, although they did not reach physical spectrum largely due to the non-closed property of the constraint algebra. In 1984, ref.\cite{Plyushchay:1984ja} analyzed the constraint structure more carefully and got deeper insight on classical solutions, but still left the full quantum spectrum undetermined.

\begin{figure}[htbp]
        \centering
        \includegraphics[width=6.5cm]{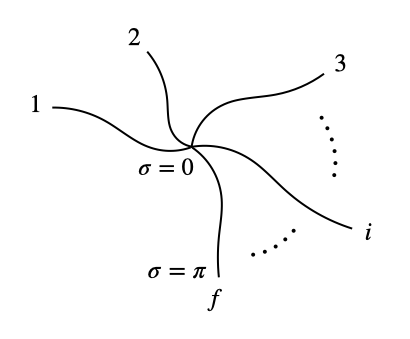}
        \caption{$f$-pronged open string junction}
        \label{fig1}
\end{figure}

In the present paper, the authors are going to revisit the problem. In particular, we propose a set of physical state conditions under which we can show there are no ghosts in the spectrum. In the following, we treat multi-pronged open bosonic string junctions in flat spacetime. One end of each string segment is tied together, whereas the other end is free (See Fig.\ref{fig1}.)
We covariantly quantize the system based on so-called ``old covariant quantization (OCQ).''

The system is different from a system of multi free strings. Instead, string segments are somehow interacting each other through the connecting point. Therefore the constraints algebra governing this system is not just a multiple of the Virasoro algebra, but rather infinite dimensional open algebra which includes a single Virasoro algebra as a subalgebra.

Also the state space is not just a multiple of single string Fock space. One needs to introduce some sets of twisted bosons in addition to a set of  ordinary bosons on the worldsheet. Accordingly, operator algebra of constraints is of twisted type, namely it contains both periodic and anti-periodic parts. Thus we find many interesting and non-trivial features worth investigating in its own right.

The paper is organized as follows. In Section~2, we formulate $f$-pronged open bosonic string junction starting from Nambu-Goto type action. We determine mode expansion of the variables in the orthonormal gauge and study structure of primary constraints. In Section~3, we investigate physical state conditions in detail. In Section~4, we concretely determine physical spectrum and discuss its properties. There are many things to be clarified part of which are discussed in the final section. Some useful algebraic relations are collected in Appnedices~A and B. Some details of the physical states are given in Appendix~C. We make a remark on light-cone gauge in Appendix~D.

\section{$f$-pronged open string junction}

Let us consider $f$-pronged open bosonic string junction shown in Fig.\ref{fig1}.
We denote a coordinate variable of $i$-th string segment by $X^{(i)\mu}(\tau,\sigma)$ $(i=1,2,\cdots,f)$, where $\mu$ $(=0,1,\cdots,D-1)$ is a space-time index and $\tau$, $\sigma$ $(\sigma\in [0,\pi])$ are the world-sheet parameters.\footnote{We adopt $\eta^{\mu\nu}=\mbox{diag.}(-1,1,\cdots,1)$ for the target space metric.} Each string is connected to the other strings at $\sigma=0$, i.e.
\begin{equation}
X^{(i)\mu}(\tau,0)=X^{(j)\mu}(\tau,0)\qquad \mbox{for arbitrary $i$ and $j$,}
\label{connect}
\end{equation}
while $\sigma=\pi$ is a free end. We also use a notation $\xi^{\alpha}=(\tau,\sigma)$ $(\alpha=0,1)$ for the world-sheet parameters. Then the Nambu-Goto type action for the system is
\begin{equation}
S=-T\sum_{i=1}^f\int d\tau d\sigma\sqrt{-\det[\partial_{\alpha}X^{(i)}{}_{\mu}(\tau,\sigma)\,\partial_{\beta}X^{(i)\mu}(\tau,\sigma)]},
\end{equation}
where $T(=\frac{1}{2\pi\alpha'})$ is the string tension and the determinant is taken with respect to the indices $\alpha,\beta$ of the partial derivatives. In the following, we sometimes abbreviate a contraction of space-time indices simply by dot; e.g.~$A\cdot B\equiv A_{\mu}B^{\mu}$. Also denote $\dot X\equiv\partial_{\tau}X$ and $X'\equiv\partial_{\sigma}X$ as a customary use. We can write the action more explicit way as
\begin{equation}
S=-T\sum_{i=1}^f\int d\tau d\sigma\sqrt{-(\dot X^{(i)}\cdot\dot X^{(i)})(X'^{(i)}\cdot X'^{(i)})+(\dot X^{(i)}\cdot X'^{(i)})^2}.\label{NGaction}
\end{equation}
The action has a reparametrization invariance under the (local) transformation
\begin{equation}
\delta X^{(i)\mu}(\tau,\sigma)=-\epsilon^{(i)\alpha}(\tau,\sigma)\,\partial_{\alpha}X^{(i)\mu}(\tau,\sigma), \label{reparametrization}
\end{equation}
as far as infinitesimal transformation parameters $\epsilon^{(i)\alpha}(\tau,\sigma)$ satisfy
\begin{equation}
\epsilon^{(i)1}(\tau,0)=0\qquad\mbox{ and }\qquad
\epsilon^{(i)1}(\tau,\pi)=0.\label{repara_bc1}
\end{equation}
Since the $i$-th term in the action only depends on the $i$-th variable $X^{(i)\mu}$, $\epsilon^{(i)\alpha}(\tau,\sigma)$ can be taken as independently of each $i$ except for the boundary condition
\begin{equation}
\epsilon^{(i)0}(\tau,0)=\epsilon^{(j)0}(\tau,0)\qquad \mbox{for arbitrary $i$ and $j$,}\label{repara_bc2}
\end{equation}
which keeps the condition (\ref{connect}).

Taking a variation $\delta X^{(i)\mu}$ in the action, we obtain a set of equations of motion 
\begin{eqnarray}
&&\hspace{-10mm}\partial_{\tau}\left(\frac{\dot X^{(i)\mu}(X'^{(i)}\cdot X'^{(i)})-X'^{(i)\mu}(\dot X^{(i)}\cdot X'^{(i)})}{\sqrt{-(\dot X^{(i)}\cdot\dot X^{(i)})(X'^{(i)}\cdot X'^{(i)})+(\dot X^{(i)}\cdot X'^{(i)})^2}}\right)\nonumber\\
&&+
\partial_{\sigma}\left(\frac{X'^{(i)\mu}(\dot X^{(i)}\cdot\dot X^{(i)})-\dot X^{(i)\mu}(\dot X^{(i)}\cdot X'^{(i)})}{\sqrt{-(\dot X^{(i)}\cdot\dot X^{(i)})(X'^{(i)}\cdot X'^{(i)})+(\dot X^{(i)}\cdot X'^{(i)})^2}}\right)
=0
\end{eqnarray}
and boundary conditions 
\begin{equation}
\left.\frac{X'^{(i)\mu}(\dot X^{(i)}\cdot\dot X^{(i)})-\dot X^{(i)\mu}(\dot X^{(i)}\cdot X'^{(i)})}{\sqrt{-(\dot X^{(i)}\cdot\dot X^{(i)})(X'^{(i)}\cdot X'^{(i)})+(\dot X^{(i)}\cdot X'^{(i)})^2}}\right|_{\sigma=\pi}=0,
\end{equation}
\begin{equation}
\left.\sum_{i=1}^f\frac{X'^{(i)\mu}(\dot X^{(i)}\cdot\dot X^{(i)})-\dot X^{(i)\mu}(\dot X^{(i)}\cdot X'^{(i)})}{\sqrt{-(\dot X^{(i)}\cdot\dot X^{(i)})(X'^{(i)}\cdot X'^{(i)})+(\dot X^{(i)}\cdot X'^{(i)})^2}}\right|_{\sigma=0}=0
\end{equation}
as a stationary condition for the action.

Canonical conjugate momentum for each variable $X^{(i)\mu}$ is
\begin{equation}
P^{(i)}_{\mu}=T\frac{\dot X^{(i)}_{\mu}(X'^{(i)}\cdot X'^{(i)})-X'^{(i)}_{\mu}(\dot X^{(i)}\cdot X'^{(i)})}{\sqrt{-(\dot X^{(i)}\cdot\dot X^{(i)})(X'^{(i)}\cdot X'^{(i)})+(\dot X^{(i)}\cdot X'^{(i)})^2}},
\end{equation}
from which we obtain primary constraints
\begin{equation}
P^{(i)}_{\mu}P^{(i)\mu}+T^2 X'^{(i)}_{\mu}X'^{(i)\mu}=0,\label{con1}
\end{equation}
\begin{equation}
P^{(i)}_{\mu}X'^{(i)\mu}=0.\label{con2}
\end{equation}
The Hamiltonian given by the Legendre transform of Lagrangian in (\ref{NGaction}) vanishes. Therefore the total Hamiltonian consists only of primary constraints multiplied by arbitrary parameter functions $u_1^{(i)}(\tau,\sigma)$ and $u_2^{(i)}(\tau,\sigma)$,
\begin{equation}
H_T=\sum_{i=1}^f\int_0^{\pi}d\sigma \left[
u_1^{(i)}(P^{(i) 2}+T^2X'^{(i) 2})+u_2^{(i)}P^{(i)}\cdot X'^{(i)}\right].
\end{equation}
The above constraints are all first-class and there are no more constraints coming from their time evolution.

Now let us take a gauge 
\begin{equation}
u_1^{(i)}=\frac{1}{2T}, \quad u_2^{(i)}=0,
\end{equation}
which is equivalent to the so-called orthonormal gauge which imposes
\begin{equation}
\dot X^{(i)}\cdot\dot X^{(i)}+X'^{(i)}\cdot X'^{(i)}=0,
\end{equation}
\begin{equation}
\dot X^{(i)}\cdot X'^{(i)}=0,
\end{equation}
or alternatively
\begin{equation}
(\dot X^{(i)}\pm X'^{(i)})^2=0.\label{orthonormal}
\end{equation}
In this gauge, the equations of motion and the boundary conditions are largely simplified as follows
\begin{equation}
\hspace{-35mm}\mbox{(Eq.~of motion)}\!\qquad (\partial_{\tau}^2-\partial_{\sigma}^2)X^{(i)\mu}=0,\label{eom}
\end{equation}
\begin{equation}
\hspace{-35mm}\mbox{(Boundary cond.)}\qquad \partial_{\sigma}X^{(i)\mu}|_{\sigma=\pi}=0,\label{bc_pi}
\end{equation}
\begin{equation}
\sum_{i=1}^f\partial_{\sigma}X^{(i)\mu}|_{\sigma=0}=0.\label{bc_0}
\end{equation}
These equations combined with (\ref{connect}) are enough to determine the mode expansion of $X^{(i)\mu}$. Note that in this gauge canonical momentum variable becomes simply $P^{(i)\mu} = T\dot X^{(i)\mu}$.
In the following, after canonically quantizing the system, we will basically impose that physical states should satisfy the relation
\begin{equation}
\langle {\rm phys} | (P^{(i)}\pm TX'^{(i)})^2 |{\rm phys}\rangle = 0.
\label{physconst}
\end{equation}
Of course, in the quantized version, some central term can appear in the constraint algebra, so we have to be careful about treating zero mode part of constraints in operator level, which will be discussed later.

\subsection{Mode expansion}
A general solution $X^{(i)\mu}(\tau,\sigma)$  of the equations of motion (\ref{eom}) consists of left-moving and right-moving modes
\begin{equation}
X^{(i)\mu}(\tau,\sigma)=X^{(i)\mu}_L(\tau+\sigma)+X^{(i)\mu}_R(\tau-\sigma).
\end{equation}
The boundary conditions also restrict the form of the functions of each mode: Eq.(\ref{bc_pi}) gives
\begin{equation}
\dot X^{(i)\mu}_L(\tau+\pi)-\dot X^{(i)\mu}_R(\tau-\pi)=0,\label{bc_pi2}
\end{equation}
while eq.(\ref{bc_0}) gives
\begin{equation}
\sum_{i=1}^f \left(\dot X^{(i)\mu}_L(\tau)-\dot X^{(i)\mu}_R(\tau)\right)=0.
\label{bc_02}
\end{equation}
And eq.(\ref{connect}) leads to
\begin{equation}
X^{(1)\mu}_L(\tau)+X^{(1)\mu}_R(\tau)=\cdots=X^{(f)\mu}_L(\tau)+X^{(f)\mu}_R(\tau),\label{connect2}
\end{equation}
which tells us that each term $X^{(i)\mu}_L(\tau)+X^{(i)\mu}_R(\tau)$ equals to a certain $i$-independent function $\phi^{\mu}(\tau)$.
Thus,
\begin{equation}
X^{(i)\mu}_R(\tau)=-X^{(i)\mu}_L(\tau)+\phi^{\mu}(\tau).
\end{equation}
Substituting this relation into eq.(\ref{bc_02}), we have
\begin{equation}
2\sum_{i=1}^f \dot X^{(i)\mu}_L(\tau)=f\dot\phi^{\mu}(\tau),
\end{equation}
which means
\begin{equation}
\phi^{\mu}(\tau)=\frac{2}{f}\sum_{i=1}^f X^{(i)\mu}_L(\tau)+c^{\mu},
\end{equation}
where $c^{\mu}$ is a constant. 
Then the right-mover is completely determined by the left-mover up to the constant:
\begin{equation}
X^{(i)\mu}_R(\tau)=\sum_{j=i}^fA_{ij}X^{(j)\mu}_L(\tau)+c^{\mu}
\end{equation}
where $A_{ij}=\frac{2}{f}-\delta_{ij}$.
Also the periodicity of the left-mover is obtained through eq.(\ref{bc_pi2}) as
\begin{equation}
\dot X^{(i)\mu}_L(\tau+2\pi)=\sum_{j=1}^fA_{ij}\dot X^{(j)\mu}_L(\tau)=\dot X^{(i)\mu}_R(\tau)\label{perio}
\end{equation}

It is easy to obtain eigenvalues and eigenvectors of the $f\times f$ matrix $A=(A_{ij})$.
Writing $A=\frac{2}{f}Z-I$ with the matrix $Z$ whose every element is 1 ($Z_{ij}=1$) and the unit matrix $I$, we can first solve the eigenvalue problem for $Z$. Apparently, a vector $v_1$ whose every element is $\frac{1}{\sqrt{f}}$ is a normalized eigenvector of $Z$ with eigenvalue $f$, and $f-1$ vectors $v_a$ ($a=2,\cdots,f$) orthogonal to $v_1$ and themselves (i.e. $v_1\cdot v_a=0$, $v_a\cdot v_b=\delta_{ab}$) are those with eigenvalue 0. Therefore, they are also the eigenvectors for $A$ as 
\begin{equation}
Av_1=v_1, \qquad Av_a=-v_a\quad(a=2,\cdots,f).
\end{equation}
The eigenvectors $v_i$ are determined by 
\begin{equation}
v_1^T=\left( \frac{1}{\sqrt{f}}\, \cdots\, \frac{1}{\sqrt{f}} \right),
\qquad
v_1\cdot v_a=0,\quad v_a\cdot v_b=\delta_{ab}.
\end{equation}
Note that one simple representation for $v_a$ ($a=2,\cdots, f$) can be chosen as follows
\begin{equation}
(v_a)_i =\left\{
\begin{array}{ll}
\frac{1}{\sqrt{a(a-1)}} &\quad (i=1,\cdots a-1)   \\
-\sqrt{\frac{a-1}{a}} & \quad (i=a)   \\
0 &\quad  (i=a+1,\cdots, f).
\end{array}
\right.\label{repv_a}
\end{equation}
By defining a real orthogonal matrix $U$ whose $i$-th row is just $v_i$ ($i=1,\cdots,f$), i.e., $U_{ij}=(v_i)_j$, diagonalization of $A$ is expressed as
\begin{equation}
UAU^T=\Gamma\equiv
\left(
\begin{array}{cccc}
 1&  &   &\\
  &-1&   &\\
  &  &\ddots&\\
  &  &   & -1
\end{array}
\right).\label{diag}
\end{equation}

Now let us define
\begin{equation}
Y^{i\mu}(\tau)\equiv\sum_{j=1}^fU_{ij}X^{(j)\mu}_L(\tau).
\end{equation}
Conversely, 
\begin{equation}
X^{(i)\mu}_L(\tau)=\sum_{j=1}^fU_{ji}Y^{j\mu}(\tau).
\end{equation}
Then from eq.(\ref{perio}),
\begin{equation}
\dot Y^{i\mu}(\tau+2\pi)=\left\{
\begin{array}{ll}
\dot Y^{1\mu}(\tau) & (i=1)   \\
-\dot Y^{i\mu}(\tau) & (i\ne 1).
\end{array}
\right.\label{perioY}
\end{equation}
If we define also for the right-mover
\begin{equation}
\tilde Y^{i\mu}(\tau)\equiv\sum_{j=1}^fU_{ij}X^{(j)\mu}_R(\tau),
\end{equation}
then
\begin{equation}
\tilde Y^{i\mu}(\tau)=\left\{
\begin{array}{ll}
 Y^{1\mu}(\tau)+\tilde c^{\mu} & (i=1)   \\
- Y^{i\mu}(\tau) & (i\ne 1)   
\end{array}
\right.
\end{equation}
where
\begin{equation}
\tilde c^{\mu}=c^{\mu}\sum_{i=1}^fU_{1i}=c^{\mu}\sqrt{f}.
\end{equation}

Now we see from eq.(\ref{perioY}) that $Y^{1\mu}(\tau)$ is periodic (up to constant) with periodicity $2\pi$ whereas $Y^{a\mu}(\tau)$ ($a=2,\cdots,f$) are anti-periodic. Therefore their mode expansions become
\begin{equation}
Y^{1\mu}(\tau)=\frac{1}{\sqrt{2}}\left[q^{\mu}+p^{\mu}\tau+i\sum_{n\ne0}\frac{\alpha^{\mu}_n}{n}e^{-in\tau}\right],
\end{equation}
\begin{equation}
\hspace{9.5mm}Y^{a\mu}(\tau)=\frac{i}{\sqrt{2}}\sum_{r\in {\bf Z}+\frac{1}{2}}
\frac{\alpha^{a\mu}_r}{r}e^{-ir\tau}\qquad \mbox{($a=2,\cdots,f$).}
\end{equation}
Bringing them all back to the original variables, we obtain (also making $q^{\mu}$ absorb a constant ambiguity $c^{\mu}$)
\begin{eqnarray}
X^{(i)\mu}(\tau,\sigma)&=&\sqrt{\frac{2}{f}}\left[q^{\mu}+p^{\mu}\tau
+i\sum_{n\ne0}\frac{\alpha^{\mu}_n}{n}e^{-in\tau}\cos(n\sigma)\right]\nonumber\\
&&+\sqrt{2}\sum_{a=2}^f(v_a)_i\sum_{r\in{\bf Z}+\frac{1}{2}}
\frac{\alpha^{a\mu}_r}{r}e^{-ir\tau}\sin(r\sigma).
\end{eqnarray}
One can easily check that this expression satisfies the equations of motion (\ref{eom}) and all the boundary conditions (\ref{connect}), (\ref{bc_pi}) and (\ref{bc_0}) by noting the property $\sum_{i=1}^f (v_a)_i=0$ which comes from $v_a\cdot v_1=0$.

To quantize the system, we will set canonical equal-time commutation relation in the interval $\sigma, \sigma' \in [0,\pi]$ with the canonical conjugate momentum $P^{(i)\mu}(\tau,\sigma)=\frac{1}{2\pi}\dot X^{(i)\mu}(\tau,\sigma)$,
\begin{equation}
\left[X^{(i)\mu}(\tau,\sigma) , P^{(j)\nu}(\tau,\sigma')\right]
=i\eta^{\mu\nu}\delta_{ij}\delta(\sigma-\sigma').
\end{equation}
Here (and hereafter) we have taken $T=\frac{1}{2\pi}$ ($\alpha'=1$) for brevity.
This relation leads to
\begin{equation}
[\alpha^{\mu}_n , \alpha^{\nu}_m]=n\delta_{n+m,0}\eta^{\mu\nu},
\qquad
[\alpha^{a\mu}_r , \alpha^{b\nu}_s]=r\delta_{r+s,0}\delta^{ab}\eta^{\mu\nu},
\qquad
[q^{\mu},p^{\nu}]=i\eta^{\mu\nu},
\label{commrelalpha}
\end{equation}
where $n,m\in{\bf Z}$,$\,$ $r,s\in{\bf Z}+\frac{1}{2}$ and $\alpha^{\mu}_0\equiv p^{\mu}$. 
Thus our total Fock space is described by $D$ ordinary bosons and $(f-1)D$ twisted bosons.

\subsection{Constraints}
Let us turn to the constraints (\ref{con1}) and (\ref{con2}).
They can be combined into
\begin{equation}
(2\pi P^{(i)}\pm X'^{(i)})^2\approx0,
\end{equation}
where wavy equal is used in the sense of eq.(\ref{physconst}).
Each of these can be written in terms of the left- and right-mover
\begin{equation}
\left(\dot X^{(i)}_L(\tau)\right)^2\approx0,\qquad \left(\dot X^{(i)}_R(\tau)\right)^2\approx0.
\end{equation}
Furthermore, in terms of the mode-diagonal variables $Y^{i\mu}$, they become
\begin{equation}
\sum_{j,k=1}^f K^i_{jk}\,\dot Y^{j}(\tau)\cdot\dot Y^{k}(\tau)\approx0,\qquad
\sum_{j,k=1}^f\tilde K^i_{jk}\,\dot Y^{j}(\tau)\cdot\dot Y^{k}(\tau)\approx0,
\label{constraintsKK}
\end{equation}
\begin{equation}
 K^i_{jk}\equiv U_{ji}U_{ki},\qquad \tilde K^i\equiv\Gamma K^i \Gamma.
\end{equation}
Here we used matrix notation $(K^i)_{jk}\equiv K^i_{jk}$ and $\Gamma$ was defined in (\ref{diag}). The constraints defined by $K^i$ and those of $\tilde K^i$ are interchanged when $\tau$ goes to $\tau+2\pi$ since $\dot{Y}^{i\mu}(\tau+2\pi)=(\Gamma \dot{Y})^{i\mu}(\tau)$. If we recombine them into those of $K^i+\tilde K^i$ and $K^i-\tilde K^i$, then each set of constraints has definite periodicity. So one may think that there are $f$ periodic and $f$ anti-periodic independent constraints. The situation, however, is not so simple.
To see this, let us look at the structure of $K^i$.

We first define the following three types of $f\times f$ symmetric matrices,
\begin{equation}
P=\left(
  \begin{tabular}{c|ccc}
$\frac{1}{f}$&  & & \\
    \hline
  &  &  &  \\ 
  &  &  &  \\ 
  &  &  &  
  \end{tabular}
\right),
\qquad
Q^{ij}=\left(
  \begin{tabular}{c|ccc}
  &  & & \\
    \hline
  &  &  &  \\ 
  &  & $\frac{1}{2}(v_a)_i(v_b)_j+\frac{1}{2}(v_a)_j(v_b)_i$ &  \\ 
  &  &  &  
  \end{tabular}
\right),
\label{defPQij}
\end{equation}
\begin{equation}
R^i=\left(
  \begin{tabular}{c|ccc}
   &  & $\frac{1}{\sqrt{f}}(v_b)_i$ & \\
    \hline
    & & &\\
 $\frac{1}{\sqrt{f}}(v_a)_i$ &  &  &  \\ 
    &  &  &
  \end{tabular}
\right).
\label{defRi}
\end{equation}
In other words, $(P)_{11}=\frac{1}{f}$, $(Q^{ij})_{ab}=\frac{1}{2}(v_a)_i(v_b)_j+\frac{1}{2}(v_a)_j(v_b)_i$, $(R^i)_{1a}=(R^i)_{a1}=\frac{1}{\sqrt{f}}(v_a)_i$ and all other elements are vanishing.
Then $K^i$ and $\tilde K^i$ can be decomposed as
\begin{equation}
K^i=P+Q^{ii}+R^i,\qquad \tilde K^i=P+Q^{ii}-R^i.
\end{equation}
One can see that matrices for anti-periodic constraints $K^i-\tilde K^i=2R^i$ are not totally independent because of the relation $\sum_{i=1}^f R^i=0$ as a result of $\sum_{i=1}^f (v_a)_i=0$. We may choose $R^a$, for example, as independent ones. Thus, we have $f$ periodic and $f-1$ anti-periodic constraints:
\begin{equation}
\sum_{j,k=1}^f (P+Q^{ii})_{jk}\,\dot Y^{j}(\tau)\cdot\dot Y^{k}(\tau)\approx0, \qquad
\sum_{j,k=1}^f R^a_{jk}\,\dot Y^{j}(\tau)\cdot\dot Y^{k}(\tau)\approx0.
\label{constraintsPQR}
\end{equation}

In order to investigate constraints in more detail, we introduce so-called Fubini-Veneziano fields $\varphi^{i\mu}(z)$ with a complex variable $z(=e^{i\tau})$.
\begin{equation}
\hspace{-12mm}\varphi^{1\mu}(z)\equiv\sqrt{2}\,Y^{1\mu}(-i\ln z)=q^{\mu}-ip^{\mu}\ln z+i\sum_{n\ne0}\frac{\alpha^{\mu}_n}{n}z^{-n},
\end{equation}
\begin{equation}
\varphi^{a\mu}(z)\equiv\sqrt{2}\,Y^{a\mu}(-i\ln z)=i\sum_{r\in{\bf Z}+\frac{1}{2}}\frac{\alpha^{a\mu}_r}{r}z^{-r}.\qquad (a=2,\cdots,f)
\end{equation}
They satisfy the periodicity 
\begin{equation}
\varphi^{1\mu}(e^{2\pi i}z)=\varphi^{1\mu}(z)+2\pi p^{\mu},
\qquad\varphi^{a\mu}(e^{2\pi i}z)=-\varphi^{a\mu}(z).
\end{equation}
It is convenient to define $A^{i\mu}(z)\equiv i\partial_z\varphi^{i\mu}(z)$, i.e.,
\begin{equation}
A^{1\mu}(z)=\sum_{n\in{\bf Z}}\alpha^{\mu}_n\,z^{-n-1},\qquad
A^{a\mu}(z)=\sum_{r\in{\bf Z}+\frac{1}{2}}\alpha^{a\mu}_r\,z^{-r-1}.
\label{defAi}
\end{equation}
$A^{1\mu}(z)$ is periodic whereas $A^{a\mu}(z)$ are anti-periodic under the replacement $z\rightarrow e^{2\pi i}z$.
If we define the operator $T_M(z)$ for a given matrix $M$ as%
\footnote{Note that the term $\frac{D}{16 z^2} {\rm tr}(M P_-)$ in the definition of $T_M(z)$  is added in order that the OPE relations can be written in a unified manner. (See \ref{app0}.)}
\begin{equation}
T_M(z)\equiv \frac{1}{2}\sum_{i,j=1}^f M_{ij}:A^{i}(z)\cdot A^{j}(z): 
\; + \;\frac{D}{16 z^2} {\rm tr}(M P_-) ,
\label{TMz}
\end{equation}
where $:{\cal O}:$ is a normal order of ${\cal O}$ with respect to the oscillators $\alpha^{\mu}_n$, $\alpha^{a\mu}_r$ (and $P_- = \frac{1}{2}(1-\Gamma)$), then 
the operators corresponding to the primary constraints eqs.(\ref{constraintsPQR}) are given by $T_{P+Q^{ii}}(z)$ and $T_{R^a}(z)$. 
These operators satisfy the relations 
\begin{equation}
T_{P+Q^{ii}}(e^{2\pi i}z)=T_{P+Q^{ii}}(z),\qquad 
T_{R^a}(e^{2\pi i}z)=-T_{R^a}(z),
\end{equation}
and they have a formal Laurent expansion with integer power of $z$ and half odd integer power respectively,
\begin{equation}
T_{P+Q^{ii}}(z)=\sum_{n\in{\bf Z}}L^{P+Q^{ii}}_n z^{-n-2},
\end{equation}
\begin{equation}
T_{R^a}(z)=\sum_{r\in{\bf Z}+\frac{1}{2}}L^{R^a}_r z^{-r-2}.
\end{equation}
In terms of the oscillators, each mode operator is written as follows
\begin{equation}
L^{P+Q^{ii}}_n=\frac{1}{2f}\sum_{m\in{\bf Z}}:\alpha_{n-m}\cdot\alpha_{m}: + \frac{1}{2}\sum_{a,b=2}^f(v_a)_i(v_b)_i\sum_{s\in{\bf Z}+\frac{1}{2}}:\alpha^a_{n-s}\cdot\alpha^b_{s}:
+\frac{D}{16} \frac{f-1}{f}\delta_{n,0},
\end{equation}
\begin{equation}
\hspace{-40mm}L^{R^a}_r=\frac{1}{\sqrt{f}}\sum_{b=2}^f(v_b)_a\sum_{m\in{\bf Z}}:\alpha^b_{r-m}\cdot\alpha_{m}:.
\end{equation}
Among the above operators, one special combination is
\begin{eqnarray}
V_n&\equiv& \sum_iL^{P+Q^{ii}}_n\\
&=&\frac{1}{2}\sum_{m\in{\bf Z}}:\alpha_{n-m}\cdot\alpha_m:+\frac{1}{2}\sum_{a=2}^f\sum_{s\in{\bf Z}+\frac{1}{2}} :\alpha^a_{n-s}\cdot\alpha^a_s:+\frac{D}{16}(f-1)\delta_{n,0}.
\end{eqnarray}
They satisfy Virasoro algebra with central charge $Df$
\begin{equation}
[V_n\,,V_m]=(n-m)V_{n+m}+\frac{Df}{12}n(n^2-1)\delta_{n+m,0}.
\end{equation}
Thus we have a single Virasoro algebra as a closed subalgebra of our constraints.


\section{Physical state condition for open string junction}

\subsection{Preliminary discussion}
For the physical state condition, it is possible to impose that the positive modes of the constraint operators should annihilate physical states just as in the case of the old covariant quantization for string:
\begin{eqnarray}
L^{P+Q^{ii}}_n |{\rm phys}\rangle &=&0,\qquad(n>0,\quad i=1,\cdots,f)
\label{physcondPQii}
\\
L^{R^a}_r |{\rm phys}\rangle&=&0.  \qquad\,(r>0,\quad a=2,\cdots,f)
\label{physcondRa}
\end{eqnarray}
Note that the algebras of $L^{Q^{i-1\,i-1}-Q^{ii}}_n$'s and $L^{R^a}_r$'s are not closed among themselves, as will be seen in Appendix~B. However, if $A|{\rm phys}\rangle=0$ and $B|{\rm phys}\rangle=0$, then $[A,B]|{\rm phys}\rangle=0$, so that requiring the above conditions will be sufficient.
On the other hand, those of zero-mode operators $L^{Q^{ii}}_0$ must be chosen more carefully.
It is not straightforward to choose appropriate conditions for general $f$ since the set of $f-1$ operators $L^{Q^{i-1\,i-1}-Q^{ii}}_0$ is not closed. 
We first study the special case of $f=2$, and go into the general discussion afterward.

\subsection{$f$=2 case}
We consider the $f=2$ case where the physical object is not a junction but an open string. 
In fact, in this case, there are only one positive integer and half-integer modes oscillators $\alpha_{n}^\mu$'s and $\alpha_{r}^{a=2, \mu}$'s, and the physical state conditions (\ref{physcondPQii}) and (\ref{physcondRa}) are reduced to  the following two simple ones
\begin{eqnarray}
V_n |{\rm phys}\rangle &=&   0,\qquad(n>0)
\\
L^{R^{a=2}}_r |{\rm phys}\rangle&=&
\left( \frac{1}{2} \sum_{m\in{\bf Z}}:\alpha_{r-m}^{a=2}\cdot\alpha_m: \right) |{\rm phys}\rangle= 0  \qquad\,(r>0)
\end{eqnarray}
since $P+Q^{11}=P+Q^{22}$. 
Also, there is only one zero mode operator $V_0$ which counts the level of the state (plus $\frac{1}{2}p^2$ and a constant) as usual for strings, and the zero mode condition should be taken as 
$V_0 |{\rm phys}\rangle = (a_0 + \frac{D}{16})  |{\rm phys}\rangle$ with a normal order constant or an intercept parameter $a_0$.
If we replace $\alpha_{n}^\mu \rightarrow \frac{1}{\sqrt{2}}\tilde\alpha_{2n}^\mu$ and 
$\alpha_{r}^{a=2, \mu} \rightarrow \frac{1}{\sqrt{2}}\tilde\alpha_{2r}^\mu$,
they satisfy $[\tilde\alpha_n^{\mu},\tilde\alpha_m^{\nu}]=n\delta_{n+m,0}\eta^{\mu\nu}$.
Writing $\tilde L_{2n}=2(V_{n}-\frac{D}{16}\delta_{n,0})$ and $\tilde L_{2n+1}=2L^{R^{a=1}}_{n+\frac{1}{2}}=-2L^{R^{a=2}}_{n+\frac{1}{2}}$, we can easily check that $\tilde L_n$'s satisfy the Virasoro algebra with central charge $D$, and
the physical state condition can be collected as 
\begin{equation}
\left(\tilde{L}_n - \tilde{a} \delta_{n,0}\right) |{\rm phys}\rangle =0 \qquad(n\ge0)
\end{equation}
where $\tilde{a}=2 a_0$ and 
\begin{equation}
\tilde{L}_n = \frac{1}{2}\sum_{m\in{\bf Z}}:\tilde\alpha_{n-m}\cdot \tilde\alpha_m:.
\end{equation}
This is exactly the same condition as for the old covariant quantization of open string theory.
As is well known, the constant $\tilde{a}$ should be taken as $\tilde{a} \le1$ and the dimension as $D\le 26$ in order to ensure that there is no ghost (negative norm) state in the physical spectrum (See~\cite{Green:1987sp} for example.)
From the various discussions like the 1-loop level unitarity or the modern BRS quantization, we know that the consistent choice should be $\tilde{a}=1$ and $D=26$.


\subsection{Zero mode condition}
Now we deal with the general $f\ge 3$ case and discuss the physical spectrum and its properties.

We have already proposed reasonable condition for positive modes of the constraint operators as (\ref{physcondPQii}) and (\ref{physcondRa}). 
Here we consider the remaining discussion on zero modes of the constraint operators given by
\begin{equation}
L^{P+Q^{ii}}_0=\frac{1}{2f}\sum_{m\in{\bf Z}}:\alpha_{-m}\cdot\alpha_{m}: + \frac{1}{2}\sum_{a,b=2}^f (v_a)_i(v_b)_i\sum_{s\in{\bf Z}+\frac{1}{2}}:\alpha^a_{-s}\cdot\alpha^b_{s}:
+\frac{D}{16} \frac{f-1}{f}.
\end{equation}
These $f$ operators are all independent and the commutation relations are 
\begin{eqnarray}
[L^{P+Q^{ii}}_0,L^{P+Q^{jj}}_0] & =& [L^{Q^{ii}}_0,L^{Q^{jj}}_0] 
\nonumber\\
&=& -\frac{1}{f} \sum_{a,b=2}^f (v_a)_{i}(v_b)_{j}
 \sum_{r \in{\bf Z}+\frac{1}{2}} r : \alpha_{-r}^a \cdot  \alpha_{r}^b:\;,
 \end{eqnarray}
which shows that the algebra is not closed within the $f$ operators $L^{P+Q^{ii}}_0$. Thus we cannot impose a condition such that a state $ |\phi \rangle $ should be a vector in the space of some finite representation of $L^{P+Q^{ii}}_0$. 
In fact, in order to obtain a closed algebra including all $f$ operators $L^{P+Q^{ii}}_0$, we have to involve infinite number of additional operators. 
On the other hand, the operator $V_0=\sum_{i=1}^f L^{P+Q^{ii}}_0$ in itself has a desirable property
\begin{equation}
V_0 |\phi \rangle = \left( N_{\rm level} +\frac{1}{2} p^2   +\frac{D(f-1)}{16} \right)|\phi \rangle 
\end{equation}
where $|\phi \rangle$ is a state whose level and the squared of the momentum are $N_{\rm level} $ and $p^2$ respectively. (Note that we take $\alpha'=1$.)
We see that one reasonable physical state condition involving the zero mode operators is 
\begin{equation}
V_0 |{\rm phys} \rangle =\left(a_0 + \frac{D(f-1)}{16} \right)|{\rm phys}  \rangle 
\label{V0a0}
\end{equation}
with some constant $a_0$.
In addition, for other zero mode operators, it is at least possible to impose the condition that $L_0^{P+Q^{ii}} |\phi \rangle $ is always physical if $|\phi \rangle $ is physical.

\subsection{Physical state condition}
\label{sec32}
From the above discussion, the most reasonable physical state condition for the present system is to impose the following set of conditions (I), (II) and (III) for arbitrary non-negative integer $N$,  positive integer $n$ and positive half-integer $r$: 
\begin{eqnarray}
&{\rm (I)} & \; L^{P+Q^{ii}}_n \left(\prod_{k\in \{j_1,\cdots, j_N \}} \!\!\!\! L^{P+Q^{kk}}_0\right) |{\rm phys}\rangle =0,\qquad(n>0, \quad i=1,\cdots,f)
\\
&{\rm (II)} &\; L^{R^a}_r \left( \prod_{k\in \{j_1,\cdots, j_N \}} \!\!\!\! L^{P+Q^{kk}}_0 \right) |{\rm phys}\rangle=0,  \qquad\,(r>0,\quad a=2,\cdots,f )
\\
&{\rm (III)} &\; V_0 |{\rm phys} \rangle =\left(a_0 + \frac{D(f-1)}{16} \right)|{\rm phys}  \rangle 
\;\;
\left(
\Leftrightarrow  \left( N_{\rm level} +\frac{1}{2} p^2   -a_0 \right) |{\rm phys} \rangle =0
\right).
\end{eqnarray}
In the above equations, 
$\prod_{k\in \{j_1,\cdots, j_N \}} L^{P+Q^{kk}}_0$ is meant to be the product of $N$ $L^{P+Q^{kk}}_0$'s :
\begin{equation}
\prod_{k\in \{j_1,\cdots, j_N \}} \!\!\!\! L^{P+Q^{kk}}_0 
= L^{P+Q^{j_1j_1}}_0 L^{P+Q^{j_2j_2}}_0 \cdots L^{P+Q^{j_ N j_N}}_0 
\end{equation}
for positive $N$, and is to be 1 for $N=0$.
In what follows, assuming that the physical state condition is indeed given by the set of these conditions, we will identify the space of physical states.

First, we examine the properties of each of the above three conditions in detail.
The simplest condition (III) determines the relation between the level and the mass of the states. 
This restriction does not affect the other two conditions since 
\begin{equation}
[V_0,L^{P+Q^{ii}}_m]= -m L^{P+Q^{ii}}_m, \qquad [V_0,L^{R^a}_r] = -r L^{R^a}_r
\end{equation}
hold for any integer $m$ or half-integer $r$. Thus we can proceed the discussion  within the space of fixed level (and mass) states separately. Now we consider the conditions (I) and (II) for $N=0$, i.e., the conditions $ L^{P+Q^{ii}}_n |{\rm phys}\rangle =0$ and $L^{R^a}_r |{\rm phys}\rangle=0$ .
In this case, from the commutation relations (\ref{CRLPQLPQ}), (\ref{CRLRLR}) and (\ref{CRLQLPR}),
we can prove that if we assume that  (I) of $n=1,2$ and (II) of $r=\frac{1}{2}, \frac{3}{2}$ for $N=0$ are satisfied, all the other (I) and (II) for $N=0$ are satisfied.
In the process of this discussion, we can also show that a state $|\phi_0\rangle$ satisfying (I) and (II) for $N=0$ always satisfies $L_n^P |\phi_0\rangle = L_n^{Q^{ij}}|\phi_0\rangle=0$ for $n\ge 2$.
Next, we impose the conditions (I) and (II) for $N=1$ on a state $|\phi_0\rangle$ satisfying the conditions for $N=0$.
From the relations (\ref{CRLRLPQ0}) and (\ref{CRLPQLPQ0}), the additional conditions we have to impose on $|\phi_0\rangle$ are
\begin{equation}
L_r^{R^a} L_0^{P}  |\phi_0\rangle  =0,
\qquad 
 L_n^{P}  |\phi_0\rangle  =0,
 \qquad
 L_n^{Q^{ii}} L_0^{Q^{jj}}  |\phi_0\rangle  =0.
 \quad(r, n>0)
 \end{equation}
In fact, it is sufficient to impose the conditions for $r=1/2$ and $n=1, 2$ since the other equations result from them. 
By continuing the similar discussion, we conclude that the conditions (I) and (II) are rewritten by the following simpler set of conditions for any $i, j, j_k \in \{1,2,\cdots,f \}$ and  $a\in \{2,\cdots,f \}$
\begin{eqnarray}
&& L_{n}^P | \phi \rangle = L_{n}^{Q^{ij}} | \phi \rangle =0,
\qquad (n>0)
\label{cond12-1}
\\
&&
L_r^{R^a} \left(L_0^{P}\right)^k  |\phi \rangle  =0,
\qquad ( r>0, \;\; k=0,1,2, \cdots )
\label{cond12-2}
\\
&&
L_n^{Q^{ii}} \left( L_0^{Q^{j_1j_1}} L_0^{Q^{j_2j_2}} \cdots L_0^{Q^{j_N j_N}} \right)  |\phi \rangle  =0.
\qquad (n>0, \;N=0,1,2, \cdots )
\label{cond12-3}
\end{eqnarray}
In particular, since $L_{0}^P= \frac{1}{2f}p^2 +\frac{1}{f}N_{\rm level}^{\rm (e)}$, the condition eq.(\ref{cond12-2}) is reduced to the set of conditions
\begin{equation}
L_{r}^{G^{(a)},s}  |\phi \rangle  =0, \qquad \left(r>0, \; s\in{\bf Z}+\frac{1}{2},\; a=2,\cdots,f  \right)
\label{12-2a}
\end{equation}
where $L_{r}^{G^{(a)},s} =: \alpha_{s}^a \cdot \alpha_{r-s}:$.
From the relations (\ref{CRLmijL0}), (\ref{CRLrQ0}), (\ref{CRLrsQ0})and (\ref{CRQGm0}), we can show that any state  
satisfying (\ref{cond12-1}) and (\ref{12-2a})  also satisfies the remaining condition (\ref{cond12-3}).

Thus, the physical state conditions (I) and (II) are reduced to the simpler conditions (\ref{cond12-1}) and (\ref{12-2a}), which we name ${\rm (I')}$, ${\rm (I'')}$ and ${\rm (II')}$ respectively:
\begin{eqnarray}
&{\rm (I')} & 
L_{n}^P | \phi \rangle =0,
\qquad (n>0)
\nonumber\\
&{\rm (I'')} & 
L_{n}^{Q^{ij}} | \phi \rangle =0,
\qquad (n>0; \, i,j=1,\cdots,f )
\nonumber\\
&{\rm (II')} &\; 
L_{r}^{G^{(a)},s}  |\phi \rangle  =0. \qquad (r>0, \; s\in{\bf Z}+\frac{1}{2},\; a=2,\cdots,f  )
\nonumber
\end{eqnarray}
In fact, it is sufficient to impose ${\rm (I')}$ for $n=1,2$, ${\rm (I'')}$ for $n=1$, ${\rm (II')}$ for $r=\frac{1}{2}$ 
since the other conditions can be derived from them.
Note that the condition ${\rm (I')}$ and ${\rm (II')}$ can be respectively represented by 
$L_{n}^E | \phi \rangle =0 $ and $L_{n}^{F^{(ab)}} | \phi \rangle =0$ $(a,b=2,\cdots,f)$.

\section{Physical spectrum and the properties}
We will explicitly solve the physical state condition given by ${\rm (I')}$, ${\rm (I'')}$, ${\rm (II')}$ and (III), and study the properties of physical spectrum. 
We represent a level $N_{\rm level}$ state as $| \phi \rangle_{\!N_{\rm level}}$ and the vacuum state that is annihilated by all the positive frequency oscillators with momentum $p$ as $|0, p \rangle$.

\subsection{First three levels: $N_{\rm level}=0, \,\frac{1}{2},\,1$}
We first identify the explicit form of physical states for $N_{\rm level}=0, \frac{1}{2},1$ and find a restriction on a constant $a_0$.

For $N_{\rm level}=0$, 
there is only one state $| \phi \rangle_{0} =|0,p \rangle$ and $p^2$ is determined by the condition (III)
as $\frac{1}{2}p^2 =a_0$.

For $N_{\rm level}=\frac{1}{2}$, a general state satisfying the mass-shell condition (III) is given by 
\begin{equation}
| \phi \rangle_{\frac{1}{2}} = \sum_{a=2}^f f^a_{\mu}\alpha^{a,\mu}_{-\frac{1}{2}} |0,p \rangle 
\end{equation}
with $\frac{1}{2} p^2= a_0 -\frac{1}{2}$.
Only the non-trivial condition for this state is given by ${\rm (II')}$ with $r=\frac{1}{2}$ and $s=\frac{1}{2}$.
This restricts the $f-1$ coefficients vectors $f^a_\mu$ to satisfy
\begin{equation}
p\cdot f^a =0.
\label{1over2cond}
\end{equation}
We see that the norm of the physical state $| f^a, p  \rangle_{\frac{1}{2}} =f^a_{\mu}\alpha^{a,\mu}_{-\frac{1}{2}} |0,p \rangle $ is 
\begin{equation}
{}_{\frac{1}{2}}\langle f^a ,p  | f^a ,p' \rangle_{\frac{1}{2}} = \frac{1}{2} f^a\!\cdot\! f^a \,\delta^D\!(p-p')
\end{equation}
where $ \langle 0 ,p  | = ( | 0 ,p \rangle)^\dagger$ with
$(\alpha^{\mu}_{n})^\dagger =\alpha^{\mu}_{-n}$,  
$(\alpha^{a,\mu}_{r})^\dagger =\alpha^{a,\mu}_{-r}$ and $ \langle 0 ,p'  | 0, p \rangle=\delta^D(p-p')$.
In this case, if we take $a_0 (=\frac{1}{2}p^2+\frac{1}{2}) > \frac{1}{2}$,  we can choose $p_0=0$ since $p^2>0$. 
Then, from the same discussion as in the case of the old covariant quantization of open string theory, $f^a$ can be taken as a time-like vector and the corresponding state $| f^a, p  \rangle_{\frac{1}{2}} $ has negative norm.
On the other hand, if we choose $a_0\le \frac{1}{2}$, physical states have always non-negative norm since $p^2 \le 0$.
In particular, for $a_0= \frac{1}{2}$, there are $f-1$ zero-norm physical states $ p_{\mu}\alpha^{a,\mu}_{-\frac{1}{2}} |0,p \rangle $ and the $(f-1)(D-2)$ transverse positive norm physical states. 
Thus, in order to ensure the no-ghost theorem for this system, we at least have to choose 
\begin{equation}
a_0\le \frac{1}{2} ,
\label{a012}
\end{equation}
which leads to
$ -\frac{1}{2} p^2\, (=\frac{1}{2} m^2 )\, \ge N_{\rm level} -\frac{1}{2}$.
If we assume this condition, all the states with level higher than $\frac{1}{2}$ has real mass ($m^2>0$). 
We later show that this assumption is plausible from the discussion of $\zeta$-function regularization calculation of normal ordering constant.

For $N_{\rm level}=1$, general states are given by 
\begin{equation}
| \phi \rangle_{1} = 
\Big( g_\mu \alpha_{-1}^\mu +
\sum_{a, b} h^{ab}_{\mu\nu} \alpha^{a,\mu}_{-\frac{1}{2}} \alpha^{b,\nu}_{-\frac{1}{2}} 
\Big) |0,p \rangle 
\end{equation}
where $h^{ab}_{\mu\nu}= h^{ba}_{\nu\mu}$ and the mass-shell condition is $\frac{1}{2} p^2= a_0 -1$.
After imposing the set of non-trivial conditions for this level, i.e., ${\rm (II')}$ with $r=\frac{1}{2}$ and $s=\pm \frac{1}{2}$ and ${\rm (I'')}$ with $n=1$, 
we obtain the general form of physical states as
\begin{equation}
| \phi_{\rm phys} ,p \rangle_{1} = \sum_{a, b} h^{ab}_{\mu\nu} \alpha^{a,\mu}_{-\frac{1}{2}} \alpha^{b,\nu}_{-\frac{1}{2}}  |0,p \rangle 
\end{equation}
where 
\begin{equation}
\eta^{\mu\nu}h^{ab}_{\mu\nu}=0, \quad p^\mu h^{ab}_{\mu\nu}=0,\quad \frac{1}{2} p^2= a_0 -1.
\label{N1cond}
\end{equation}
We see that the norm of this state is calculated as
\begin{equation}
{}_{1\!}\langle  \phi_{\rm phys},p  | \phi_{\rm phys} ,p'\rangle_{\!1} 
= \frac{1}{2}  \sum_{a, b} h^{ab}_{\mu\nu} h^{ab,\mu\nu}
\,\delta^D\!(p-p')
\label{N1norm}
\end{equation}
If we assume (\ref{a012}), mass-shell condition for this (and all the higher) level ensures $p^2<0$, and we can choose the frame $p_{\mu}=\delta_{\mu}{}^{0}p_0$ ($p_0\ne 0$ and $p_{i(\ne 0)} =0$). 
Then, from the conditions (\ref{N1cond}), $h_{\mu\nu}^{ab}=0$ if $\mu$ or $\nu$ equals to 0, and 
the norm (\ref{N1norm}) becomes positive for any non-trivial $h_{\mu\nu}^{ab}$.
Note that there is no zero-norm physical state for any $a_0\le \frac{1}{2}$.

\subsection{Physical states for an arbitrary level $N_{\rm level}$}
We now investigate the physical state condition and obtain general form of physical states for an arbitrary level $N_{\rm level}$.
Fortunately, physical state condition we set is strong enough to completely identify the space of physical states for any level.
The result is the following:
Any state satisfying the physical state condition has the form 
\begin{equation}
h^{a_1a_2\cdots a_K}_{[\mu^M_1 \mu^M_2\cdots \mu^M_f \cdots \cdots  \mu^2_1 \mu^2_2\cdots \mu^2_f\mu^1_1 \mu^1_2\cdots \mu^1_f]\sigma_1\sigma_2\cdots \sigma_K}
\left(
\prod_{m=1}^M
\alpha_{-m}^{\mu^m_1} \alpha_{-(m-1/2)}^{2,\mu^m_2} \cdots \alpha_{-(m-1/2)}^{f,\mu^m_f} \right)
\prod_{i=1}^K\alpha_{-\frac{1}{2}}^{a_i,\sigma_i}
| 0,p\rangle
\label{generalh1}
\end{equation}
or
\begin{eqnarray}
&&\hspace{-1cm}h^{a_1a_2\cdots a_K}_{[\mu^{M+1}_2\cdots \mu^{M+1}_f  \mu^M_1 \mu^M_2\cdots \mu^M_f \cdots \cdots  \mu^2_1 \mu^2_2\cdots \mu^2_f \mu^1_1 \mu^1_2\cdots \mu^1_f] \sigma_1\sigma_2\cdots \sigma_K}
\nonumber\\
&&\hspace{1cm}\times
\alpha_{-(M+1/2)}^{2,\mu^{M+1}_2} \cdots \alpha_{-(M+1/2)}^{f,\mu^{M+1}_f}
\left(
\prod_{m=1}^M
\alpha_{-m}^{\mu^m_1} \alpha_{-(m-1/2)}^{2,\mu^m_2} \cdots \alpha_{-(m-1/2)}^{f,\mu^m_f} \right)
\prod_{i=1}^K\alpha_{-\frac{1}{2}}^{a_i,\sigma_i}
| 0,p\rangle.
\label{generalh2}
\end{eqnarray}
Here, $M$ and  $K$ are non-negative integers. 
Also, $h_{[\cdots],\cdots}$ is a tensor field with all the spacetime indices $\mu_a^m$ within the bracket $[\cdots]$ are anti-symmetric.
The tensor field $h_{\cdots}$ given in eq.(\ref{generalh1}) should satisfy the relations 
\begin{eqnarray}
&&p^{\mu_a^m} h^{a_1a_2\cdots a_K}_{[\mu_1^M \mu_2^M \cdots  \mu^1_1 \mu^1_2\cdots \mu^1_f] \sigma_1\sigma_2\cdots \sigma_K} =
p^{\sigma_i}
h^{a_1a_2\cdots a_K}_{[\mu_1^M \mu_2^M \cdots  \mu^1_1 \mu^1_2\cdots \mu^1_f] \sigma_1\sigma_2\cdots \sigma_K} =0
\label{hcond1}
\\
&&\eta^{\sigma_i \mu^1_k} h^{a_1a_2\cdots a_K}_{[\mu_1^M \mu_2^M \cdots  \mu^1_1 \mu^1_2\cdots \mu^1_f] \sigma_1\sigma_2\cdots \sigma_K} =
\eta^{\sigma_i \sigma_j}
h^{a_1a_2\cdots a_K}_{[\mu_1^M \mu_2^M \cdots  \mu^1_1 \mu^1_2\cdots \mu^1_f] \sigma_1\sigma_2\cdots \sigma_K} =0,
\label{hcond2}
\\
&& h^{a_1a_2\cdots a_K}_{[\mu_1^M \mu_2^M \cdots  \mu^1_1 \mu^1_2\cdots \mu^1_f] \sigma_1\sigma_2\cdots \sigma_K} 
\quad \mbox{is symmetric under a permutation $(a_i,\sigma_i )\, \leftrightarrow \,(a_j,\sigma_j )$ }
\label{hcond3}
\end{eqnarray}
for any $i$ and $j$ ($i\ne j$).
Also, $h_{\cdots}$ given in eq.(\ref{generalh2}) satisfies the corresponding relations. 
Note that the level of the state (\ref{generalh1}) is 
\begin{equation}
\frac{K}{2}+ \sum_{m=1}^{M} \left(m+  (f-1)(m-1/2) \right)
=\frac{1}{2}(K+fM(M+1) -Mf+M ),
\end{equation}
and that of (\ref{generalh2}) is $\frac{1}{2}(K+(M+1)(Mf+f-1) )$.
The proof that all the physical states are given by the states of the form  (\ref{generalh1}) or (\ref{generalh2}) is given in Appendix~C.

As a non-trivial example, we represent all the physical states for $f=3$ and $N_{\rm level}=5$.
There are three types of physical states:
\begin{eqnarray}
&&
h_{[\mu^2_2  \mu^2_3 \mu^1_1\mu^1_2\mu^1_3 ]}
\alpha_{-\frac{3}{2}}^{2,\mu^2_2} \alpha_{-\frac{3}{2}}^{3,\mu^2_3} 
\alpha_{-1}^{\mu^1_1} \alpha_{-\frac{1}{2}}^{2,\mu^1_2} \alpha_{-\frac{1}{2}}^{3,\mu^1_3} 
| 0,p\rangle ,
\\
&&
h^{a_1a_2\cdots a_6}_{[\mu^1_1\mu^1_2\mu^1_3 ]\sigma_1\sigma_2\cdots \sigma_6}
\alpha_{-1}^{\mu^1_1} \alpha_{-\frac{1}{2}}^{2,\mu^1_2} \alpha_{-\frac{1}{2}}^{3,\mu^1_3} 
 \alpha_{-\frac{1}{2}}^{a_1,\sigma_1} \alpha_{-\frac{1}{2}}^{a_2,\sigma_2}  \cdots  \alpha_{-\frac{1}{2}}^{a_6,\sigma_6} 
| 0,p\rangle ,
\\
&&
h^{a_1a_2\cdots a_{10}}_{\sigma_1\sigma_2\cdots \sigma_{10}}
 \alpha_{-\frac{1}{2}}^{a_1,\sigma_1} \alpha_{-\frac{1}{2}}^{a_2,\sigma_2}  \cdots  \alpha_{-\frac{1}{2}}^{a_{10},\sigma_{10} }
| 0,p\rangle 
\end{eqnarray}
where $a_i =2$ or $3$.
Each field satisfies the following relations
\begin{eqnarray}
&& p^{\mu^2_2} h_{[\mu^2_2  \mu^2_3 \mu^1_1\mu^1_2\mu^1_3 ]}=0,
\\
&&
p^{\mu^1_1}  h^{a_1a_2\cdots a_6}_{[\mu^1_1\mu^1_2\mu^1_3 ]\sigma_1\sigma_2\cdots \sigma_6}
= 
p^{\sigma_i}  h^{a_1a_2\cdots a_6}_{[\mu^1_1\mu^1_2\mu^1_3 ]\sigma_1\sigma_2\cdots \sigma_6}=0,
\nonumber\\
&&\qquad
\eta^{\mu^1_1\sigma_i}  h^{a_1a_2\cdots a_6}_{[\mu^1_1\mu^1_2\mu^1_3 ]\sigma_1\sigma_2\cdots \sigma_6}
=\eta^{\sigma_i\sigma_j}  h^{a_1a_2\cdots a_6}_{[\mu^1_1\mu^1_2\mu^1_3 ]\sigma_1\sigma_2\cdots \sigma_6}=0,
\\
&&
p^{\sigma_i} h^{a_1a_2\cdots a_{10}}_{\sigma_1\sigma_2\cdots \sigma_{10}} =0,
\quad
\eta^{\sigma_i\sigma_j} 
h^{a_1a_2\cdots a_{10}}_{\sigma_1\sigma_2\cdots \sigma_{10}} =0
\end{eqnarray}
and the appropriate symmetric properties.

\subsection{Properties of general physical states}
Now that we have fully identified the possible form of the physical states, we investigate and summarize their general properties. 

First, we consider the constant $a_0$ which appears in eq.(\ref{V0a0}) and has not been determined yet. 
In the case of string theory, the corresponding constant can be identified by calculating the zero-point energy of the sum of all physical degrees of freedom and the result is confirmed by the discussion of the BRS quantization method.
For our system, the similar calculation can be performed if we assume the number of physical degrees of freedom is the same as that of the transverse degrees of freedom.
The sum of the zero-point energy corresponding to the degrees of freedom for one space-time direction for our system is given by 
\begin{equation}
 \sum_{n=0}^\infty n + (f-1) \sum_{n=0}^{\infty} \left( n+\frac{1}{2} \right)
\end{equation}
multiplied by $-\frac{1}{2}$.
This type of summation can be performed by using the $\zeta$-function regularization method and the result is given by 
\begin{equation}
  \sum_{n=0}^\infty n  \rightarrow -\frac{1}{12},
  \qquad
   \sum_{n=0}^{\infty} \left( n+\frac{1}{2} \right) \rightarrow 
  \frac{1}{24}.
\end{equation}
By using the result, 
\begin{equation}
  \sum_{n=0}^\infty n + (f-1) \sum_{n=0}^{\infty} \left( n+\frac{1}{2} \right) \rightarrow 
  \frac{f-3}{24},
  \label{zeroptcal}
\end{equation}
and if we assume the number of physical degrees of freedom is same as that of the transverse degrees of freedom, which is the case for the string theory,  $a_0$ is given by eq.(\ref{zeroptcal})
multiplied by $-\frac{D-2}{2}$ as
\begin{equation}
a_0 = - (D-2) \frac{f-3}{48}.
\label{a0lczetacalc}
\end{equation}
For any $f\ge 3$, and for the string case $f=2$ and $D=26$, this result indeed satisfies the relation $a_0\le \frac{1}{2}$ which is necessary for all level $N_{\rm level}=\frac{1}{2}$ physical states to have positive norm. 
In particular, for $D=26$, $a_0= -\frac{f-3}{2}$, and the on-shell condition (\ref{V0a0}) restricts 
\begin{equation}
-\frac{1}{2}p^2\left( = \frac{1}{2}m^2  \right)=N_{\rm level} +\frac{f-3}{2},
\label{mNrel} 
\end{equation}
 which means that all states satisfying the on-shell condition should have integer or half-integer $-\frac{1}{2}p^2$ for every $f$ including the string $f=2$ case. 
In particular, $m^2=f-3$ for $N_{\rm level}=0$ and $m^2=f-2$ for $N_{\rm level}=\frac{1}{2}$. Thus,  physical spectrum for $f\ge 3$ is limited to the massive one except for  $f=3$ and $N_{\rm level}=0$ in which case the ground state $|0, p\rangle$ is massless and physical. 

From eq.(\ref{mNrel}), we can also study the relation between mass and spin for each level. 
In general, the highest spin state for level $N$ is given by the state of the form 
\begin{equation}
h_{(\sigma_1\sigma_2\cdots \sigma_{2\!N} )}^{a_1a_2\cdots{a_{2\!N}}}
\alpha_{-\frac{1}{2}}^{a_1,\sigma_1} \alpha_{-\frac{1}{2}}^{a_2,\sigma_2}
\cdots \alpha_{-\frac{1}{2}}^{a_{2\!N},\sigma_{2\!N}} |0,p \rangle .
\end{equation}
Any other physical state for the level has spin less than that of this state:
$0\le J \le 2N(=J_{\max})$.
 Thus, for arbitrary $f\ge3$, the relation between mass $m$ and spin $J$ is represented  by
\begin{equation}
J\le m^2 -f+3\, ,
\end{equation}
which is also applied to the $f=2$ open string case. (Remember that we take $\alpha'=1$.)

Next, we will show the No-ghost theorem that ensures all the physical states have positive norm under the assumption $a_0\le \frac{1}{2}$ for general $f\ge 3$.
We have already shown the theorem is indeed met for $N_{\rm level}\le 1$ in section~4.1.
As for the discussion on general $N_{\rm level}$, we first note that any on-shell physical state has $-\frac{1}{2} p^2>0$ for $f\ge 3$ and $N_{\rm level}>0$ under the assumption $a_0\le \frac{1}{2}$. 
Then, choosing the momentum frame as $p_\mu=\eta_{\mu0} p^0$, we see that any component of the tensor field $h_{\nu_1\nu_2 \cdots }$ corresponding to any physical state of the form (\ref{generalh1}) or (\ref{generalh2}) always vanishes if any space-time index $\nu_j=0$.
That is, any non-zero physical state (\ref{generalh1}) or (\ref{generalh2}) includes only space-like oscillators $\alpha_{-n}^i$ and $\alpha_{-s}^{ai}$.
From the properties of the commutation relations of the oscillators, we easily see that all such states have positive norm.
Thus we have proven the No-ghost theorem.

From the above discussion, we see that there is no zero norm physical state unlike the case of string theory.
Zero norm physical states for string theory play an important role for the theory to be equipped with gauge symmetry.
For $f\ge 3$, there seems to be no gauge degrees of freedom in the physical spectrum.

\section{Discussion}
We have revisited the covariant quantization problem of $f$-pronged open string junction. Its excitation is described by a set of ordinary bosons as well as $f-1$ sets of twisted bosons on the world sheet. The constraints form an open algebra with operators obeying both periodic and anti-periodic boundary conditions. We have, for the first time, succeeded in giving the physical state condition and identify the physical states which indeed have positive norm.
Several remarks and/or questions are in order.

One may wonder whether our result of physical states coincides with light-cone gauge which is familiar in the ordinary single string case. This question is not so simple to be answered because taking a light-cone gauge is a non-trivial issue for string junction as described in the \ref{lightcone}. Indeed, naive truncation to the transverse oscillators gives slightly different number of states in each level. For example, in $N_{\rm level}=1$ (with $D\ge 3$, $f\ge 2$) case the number of our physical states is
\begin{equation}
\frac{1}{2}(f-1)(D-2)[(f-1)(D-2)+1]+(f-1)^2(D-2),
\end{equation}
 while naive truncation to the transverse mode gives 
\begin{equation}
\hspace{-14mm}\frac{1}{2}(f-1)(D-2)[(f-1)(D-2)+1]+(D-2).
\end{equation}
The number of constraints seems enough to eliminate light-cone degrees of freedom, but structure of physical states is not the same as naive truncation to the transverse oscillators. 

As is explained, the constraint algebra is not closed. If we include all operators newly appeared in the commutator one by one, then we eventually obtain a very large algebra. It is interesting to understand this algebra and to interpret our physical states in terms of it. In fact, by construction our physical states in each level will become some sort of representation of a zero-mode subalgebra of that. If we define 
\begin{equation}
B^{(p)}_{ab}=\sum_{r\in{\bf Z}+\frac{1}{2}}r^{p-1}:\alpha^a_{-r}\cdot\alpha^b_{r}:\qquad p=1,2,\cdots
\end{equation}
which appears in the zero mode part of the above mentioned large algebra, they satisfy
\begin{equation}
\left[ B^{(p)}_{ab}\,,\, B^{(q)}_{cd}\right]
=\delta_{bc}B^{(p+q)}_{ad}+(-1)^{p-1}\delta_{ac}B^{(p+q)}_{bd}
+(-1)^{q-1}\delta_{bd}B^{(p+q)}_{ac}+(-1)^{p+q-2}\delta_{ad}(-1)^{q-1}B^{(p+q)}_{bc},
\end{equation}
\begin{equation}
B^{(p)}_{ab}=(-1)^{p-1}B^{(p)}_{ba}.
\end{equation}
Note that $B^{(1)}_{ab}=2L_0^{F^{(ab)}}-\frac{D}{8}\delta^{ab}$ in terms of the operator defined in (\ref{defLnFab}).
This algebra is isomorphic to a twisted (${\bf Z}_{>0}$)-graded version of gl$(f-1)$, i.e., $\{E^+_{ab}\otimes u^{2k-1},E^-_{ab}\otimes u^{2k}\, |\, k=1,2,\cdots;\, E^{\pm}_{ab}=\pm E^{\pm}_{ba} \in {\rm gl}(f-1);\, u \in {\bf C}\}$.
We hope this point will be clarified in the future.

Our analysis in the present paper has been within so-called old covariant quantization (OCQ). One can find other fine structures from new covariant or BRST quantization which will be our next task. After getting it done, we are also able to construct free field theory for string junction. The authors previously studied extended string field theory~\cite{Asano:2013rka,Asano:2016rxi} where multiple string Fock spaces are utilized to describe massless higher spin modes with massive tower, just as closed string field theory can be formulated by doubled Fock space of open string with suitable matching condition. Fock space structure there is very similar to the current one, so that the adaptation  of the formalism to the string junction may be straightforward.

Another future task is to consider interactions. In \cite{Kiriyama:1976ue} single string emission vertex from the free end of a string segment was considered. There may be more varieties of interactions some of which may need introduction of other type of junction. We need to classify them anyway.

\section*{Acknowledgements}
The work was supported in part by JSPS KAKENHI Grant Number JP21K03579 (M.A.), JP20K03966 and JP21K03537 (M.K.). 
\appendix 
\def\thesection{Appendix~\Alph{section}}
\renewcommand{\theequation}{\Alph{section}.\arabic{equation}}
\setcounter{equation}{0}
\setcounter{figure}{0}

\section{$\!\!$ OPE for $T_M(z)$}

\label{app0}
\setcounter{equation}{0}
We study the OPE properties of the operators $T_M(z)$ defined by eq.(\ref{TMz}) for general $M$.
The product of two operators $A^{i\mu}(z)$ and $A^{i\mu}(w)$ is rewritten by normal-ordered form by using the commutation relations  (\ref{commrelalpha}) as 
\begin{equation}
A^{1\mu}(z) A^{1\nu} (w) =
\frac{1}{(z-w)^2} \eta^{\mu\nu} \; +\;: A^{1\mu}(z) A^{1\nu} (w) :
\end{equation}
and
\begin{equation}
A^{a\mu}(z) A^{b\nu} (w) =\left(
\frac{1}{(z-w)^2} +\epsilon(z,w) \right) \delta^{ab} \eta^{\mu\nu} \; +\; : A^{a\mu}(z) A^{b\nu} (w) :
\end{equation}
where
\begin{equation}
\epsilon(z,w) = \frac{1}{2 \sqrt{zw} ( \sqrt{z} + \sqrt{w} )^2}\;.
\end{equation}
By using these relations, the product of $T_M(z)$ and $T_N(w)$ for general $f\times f$ symmetric matrices $M$ and $N$ can be calculated as 
\begin{eqnarray}
T_M(z) T_N(w) &\sim& 
 \frac{D}{2}\frac{1}{(z-w)^4} {\rm tr}(M\circ N)
+\frac{2}{(z-w)^2} T_{M\circ N}(w) +\frac{1}{z-w} \partial_w T_{M\circ N}(w)
\nonumber\\
&&\hspace*{-1.5cm}+\frac{1}{4}\frac{1}{z-w}\left([M,N] \right)_{ij}
 \left( :  (\partial_w A^i(w)) \cdot A^{j}(w) :
-  :A^i (w)\cdot \partial_w A^{j}(w) :\right)
\label{OPETT}
\end{eqnarray}
where
\begin{equation}
M\circ N \equiv \frac{1}{2}(MN+NM).
\end{equation}
Note that in the right hand side of eq.(\ref{OPETT}), there remain terms which cannot be represented only by $T$ unless $[M,N]= 0$. 

We investigate the properties of (\ref{OPETT}) for general $M$ and $N$.
First, note that any symmetric $f\times f$ matrix $M$ can be expanded by the base matrices $H^{AB}=H^{BA}=(H^{AB})^T$ ($A, B = 1,2, \cdots,f$) 
\begin{equation}
(H^{(AB)}{})_{ij} =\frac{1}{2} \left( \delta_{iA}\delta_{jB}+  \delta_{iB}\delta_{jA}\right).
\end{equation}
We also use the following expressions
\begin{equation}
E=H^{11}, \qquad F^{(ab)} = H^{ab} ,\qquad G^{(a)}=H^{(1a)}
\end{equation}
where  $a=2,\cdots, f$.
We can divide these base matrices  $H^{(AB)}$ into two classes and define
\begin{equation}
{\cal M}^+ ={\rm Span}\{H^{11}, H^{ab}\}, \qquad {\cal M}^-={\rm Span}\{H^{1a}\} .
\end{equation}
For any matrix $M^{\pm}\in {\cal M}^{\pm}$, from the definition of  $T_M(z)$ given by eq.(\ref{TMz}), 
the following relation holds:
\begin{equation}
T_{M^\pm} (e^{2\pi i} z) = \pm T_{M^\pm} (z).
\end{equation}
The mode expansion of $T_{M^\pm}(z)$ is given by 
\begin{equation}
T_{M^+}(z)=\sum_{n\in{\bf Z}} L^{M^+}_n z^{-n-2},
\qquad
T_{M^-}(z)=\sum_{r\in{\bf Z}+\frac{1}{2}}L^{M^-}_r z^{-r-2}.
\end{equation}
Now that we have prepared the appropriate base matrices $H^{AB}$ and the mode expansion for $T_M(z)$, we can calculate the commutation relation $[L^M_\xi, L^N_\eta]$ by operating 
\begin{equation}
\oint_{C_0}\frac{dw}{2\pi i} \oint_{C_w}\frac{dz}{2\pi i} \, z^{\xi+1} w^{\eta+1} 
\end{equation}
on eq.(\ref{OPETT}) for any two matrices $M$ and $N$.
Here $C_0$ and $C_w$ represent contour integration around $w=0$ and $z=w$ respectively. 
Note that $\xi$ (or $\eta$) is an integer if the matrix $M$ (or $N$) belongs to ${\cal M}^+$, and a half-integer if $M$ (or $N$) belongs to ${\cal M}^-$.
After performing the integration calculation, we obtain the commutation relation 
\begin{eqnarray}
[L^M_\xi, L^N_\eta] &=& (\xi-\eta) L_{\xi+\eta}^{M\circ N} +
\frac{D}{12}  \mathrm{Tr}(M\circ N) (\xi^3-\xi) \delta_{\xi+\eta,0} 
\nonumber\\
&& \qquad +\frac{1}{4} \left([M,N] \right)_{ij} \sum_\zeta (2\zeta -\xi-\eta)
:\alpha_{\xi+\eta-\zeta}^i \cdot \alpha^j_{\zeta} :\;.
\label{LxiLeta}
\end{eqnarray}
Here $\zeta$ is an integer for $j=1$ and and a half-integer for $j=a$.

In the following, we explicitly present the commutation relations for each pair of base matrices $H^{(AB)}$ (or  $E$, $F^{(ab)}$ and $G^{(a)}$) after collecting the related useful relations.

The base matrices satisfy the relations
\begin{eqnarray}
H^{(AB)}\circ H^{(CD)} &=& 
\frac{1}{4}
\left(
\delta_{AC}H^{BD}+\delta_{AD}H^{BC}+\delta_{BC}H^{AD}+\delta_{BD}H^{AC}
\right),
\\
\left[ H^{(AB)}, H^{(CD)} \right]_{ij}&=& 
\frac{1}{4}
\left(
\delta_{AC} (\delta_{Bi} \delta_{Dj} -\delta_{Bj} \delta_{Di})
+\delta_{AD} (\delta_{Bi} \delta_{Cj} -\delta_{Bj} \delta_{Ci})
\right.
\nonumber\\
&&
\qquad
\left. +\delta_{BC} (\delta_{Ai} \delta_{Dj} -\delta_{Aj} \delta_{Di})
+\delta_{BD} (\delta_{Ai} \delta_{Cj} -\delta_{Aj} \delta_{Ci})
\right).
\end{eqnarray}
The mode operators for base matrices are given by
\begin{equation}
L_n^{E} = \frac{1}{2} \sum_{m\in {\bf Z}}:\alpha_{n-m}\cdot\alpha_{m}: ,
\end{equation}
\begin{equation}
L_n^{F^{(ab)}} = \frac{1}{2} \sum_{r \in {\bf Z}+\frac{1}{2}}:\alpha^{a}_{n-r}\cdot\alpha^{b}_{r}: 
 \;+\frac{D}{16} \delta^{ab} \delta^{n,0} ,
 \label{defLnFab}
\end{equation}
\begin{equation}
L_r^{G^{(a)}} = \frac{1}{2} \sum_{m \in {\bf Z}}:\alpha^{a}_{r-m}\cdot\alpha_{m}: .
\end{equation}
Commutation relations are summarized as follows.
\begin{equation}
[L_m^E, L_n^E] = (m-n) L_{m+n}^E +\delta_{m+n,0} \frac{D}{12}m(m^2-1),
\label{CRLELE}
\end{equation}
\begin{eqnarray}
[L_m^{F^{(a_1 a_2)}}, L_n^{F^{(b_1 b_2)}}]&=& 
\sum_{i=1}^2 \sum_{j=1}^2 \delta^{a_{i+1}, b_{j+1}} 
\left\{
\frac{1}{4} (m-n) L_{m+n}^{F^{a_{i}b_{j} }}
\right.
 \nonumber \\
 && 
 \qquad \qquad \qquad
\left.
+\frac{1}{8}  r
 \left(:  \alpha^{a_i}_{m+n-r}\cdot  \alpha^{b_j}_{r} : - : \alpha^{b_j}_{m+n-r}\cdot  \alpha^{a_i}_{r}   :\right)
\right\}
 \nonumber \\
 &&
 + \;\delta_{m+n,0} \left(   \delta^{a_{1}, b_{1}}    \delta^{a_{2}, b_{2}}  +  \delta^{a_{1}, b_{2}}    \delta^{a_{2}, b_{1}}   \right)   
 \frac{D}{24} 
 m \left(m^2-1\right) ,
\label{relFF}
\end{eqnarray}
\begin{eqnarray}
[L_r^{G^{(a)}}, L_s^{G^{(b)}}] &=& \frac{1}{4} (r-s) \left( L_{r+s}^{F^{(ab)}} + \delta^{a,b}  L_{r+s}^E \right)
 +\frac{1}{8} \sum_{t \in  {\bf Z} +\frac{1}{2}}t 
\left(:  \alpha^a_{r+s-t}\cdot  \alpha^b_{t} : - : \alpha^b_{r+s-t}\cdot  \alpha^a_{t}   :\right)
\nonumber\\
&&
\qquad +\delta_{r+s,0}\delta^{ab} \frac{D}{24}r \left(r^2-1\right) ,
\end{eqnarray}
\begin{equation}
 [L_n^E,  L_s^{G^{(a)}} ]
=
 \frac{1}{2} \sum_{m \in  {\bf Z} }  (n-m) : \alpha^{a}_{s+n-m}  \cdot   \alpha_m :\,,\qquad\qquad
 \end{equation}
\begin{eqnarray}
[L_n^{F^{(ab)}}, L_s^{G^{(c)}}] &=& \delta^{a,c}\,\frac{1}{4} \sum_{m \in  {\bf Z} }  (m-s) : \alpha^{b}_{s+n-m}  \cdot   \alpha_m :
\nonumber \\
&&
+ 
 \delta^{b,c}\,\frac{1}{4} \sum_{m \in  {\bf Z} }  (m-s) : \alpha^{a}_{s+n-m}  \cdot   \alpha_m : \;.
 \label{relFG}
\end{eqnarray}
Note that  $i$ and $j$ are taken as mod 2 values in eq.(\ref{relFF}).
We see that the operators $L_n^E$ form the Virasoro algebra with central charge $D$.
On the other hand, the set of all the operators  $L_n^E$, $L_n^{F^{(ab)}}$ and $L_r^{G^{(a)}}$ does not form a closed algebra because of the last term of eq.(\ref{LxiLeta}).

Commutation relations for general matrices are obtained from the above relations by using  
\begin{equation}
L_\xi^{M+ N} = L_\xi^{M} + L_\xi^{N}
\end{equation}
which follows from 
\begin{equation}
T_M (z) + T_N (z) =T_{M+N} (z) .
\end{equation}


\section{$\!\!$ Algebra of $L_\xi^{M}$ and the related useful relations}

\label{app1}
\setcounter{equation}{0}
We further investigate the properties of the algebra given by the operators $L_\xi^{M}$ obtained by the mode expansion of $T_M(z)$ especially for $M=P, Q^{ij}, R^i$.
The matrices $P$, $Q^{ij}$ and $R^i$ defined in eqs.(\ref{defPQij}) and (\ref{defRi}) are expanded by the base matrices $E$, $F^{(ab)}$ and $G^{(a)}$ as
\begin{eqnarray}
P &=& \frac{1}{f}E, \\
 Q^{ij} &=& \sum_{a=2}^f \sum_{b=2}^f  (v_a)_i  (v_b)_j F^{(ab)},\\
 R^i &=& \sum_{a=2}^f\frac{2}{\sqrt{f}} (v_a)_i G^{(a)}.
\end{eqnarray} 
We see that the relations 
\begin{equation}
\sum_{i=1}^f Q^{ij}=0, \quad \sum_{i=1}^f R^{i}=0
\end{equation}
hold since 
\begin{equation}
\sum_{i=1}^f (v_a)_i =0, \qquad  \sum_{i=1}^f (v_a)_i (v_b)_i=\delta_{ab} .
\end{equation}
Note that the following relation is also useful.
\begin{equation}
\sum_{a=2}^f (v_a)_i (v_a)_j = \delta_{ij}-\frac{1}{f}  \;\left( \;= {\rm Tr}\,Q^{ij} \;\right)\;.
\end{equation}

From the general formula of commutation relations (\ref{LxiLeta}), We can derive the following set of relations that are useful for analyzing the physical state conditions.
\begin{eqnarray}
 \frac{1}{2} [L_m^{Q^{ii}}, L_n^{Q^{jj}}]
+ 
\frac{1}{2} [L_m^{Q^{jj}}, L_n^{Q^{ii}}]
&=&
 (m-n) 
\left(\delta_{ij}-\frac{1}{f} \right) L_{m+n}^{Q^{ij}} 
   \nonumber\\
   &&
   \hspace*{-2cm}
  + \,\delta_{m+n,0} \left(  
   \delta^{ij} -\frac{1}{f}  \right)^2   
 \frac{D}{12} 
 m \left(m^2-1\right) ,
  \label{CRLPQLPQ}
\end{eqnarray}
\begin{eqnarray}
&&\hspace*{-1cm} \frac{1}{2} [L_r^{R^i}, L_s^{R^j}] +\frac{1}{2}  [L_r^{R^j}, L_s^{R^i}] 
= (r-s) \left( 
\frac{1}{f} L_{r+s}^{Q^{ij}} +\left( \delta_{ij} -\frac{1}{f}\right) 
L_{r+s}^P
\right)
\nonumber\\
&& \hspace*{7cm}+\delta_{r+s,0} \frac{1}{f}\left( \delta_{ij} -\frac{1}{f}\right) 
\frac{D}{6}r \left(r^2-1\right) ,
 \label{CRLRLR} 
\end{eqnarray}
\begin{eqnarray}
&&\hspace*{-1cm} \frac{1}{f} [L_n^{Q^{ij}}, L_s^{R^{k}}] 
+ \frac{1}{2}\left( \delta_{jk} -\frac{1}{f}\right) [L_n^P,  L_s^{R^{i}} ]
+ \frac{1}{2}\left( \delta_{ik} -\frac{1}{f}\right) [L_n^P,  L_s^{R^{j}} ]
\nonumber\\
&&  \hspace*{3cm}=
 \frac{1}{2f} (n-s)  \left\{
 \left( \delta_{jk} -\frac{1}{f}\right) L_{n+s}^{R^i}  + \left( \delta_{ik} -\frac{1}{f}\right) L_{n+s}^{R^j} 
\right\} .
 \label{CRLQLPR}
\end{eqnarray}
Also, note that 
\begin{equation}
[L^{P}_m\,,L^{P}_n]=(m-n)\frac{1}{f} L^{P}_{m+n} + \delta_{m+n,0} \frac{D}{12 f^2}m(m^2-1)  .
\end{equation}

Next, we give several useful commutation relations between zero mode operators and non-zero mode operators:
\begin{eqnarray}
 [L_r^{R^j}, L_0^{P+Q^{ii}}] 
= \left( \delta_{ij} -\frac{1}{f}\right) r L_r^{R^i}
+\frac{1}{2} 
\left(2- \delta_{ij} f\right)
 [L_r^{R^i} +L_r^{R^j} , L_0^{P}] ,
 \label{CRLRLPQ0} 
\end{eqnarray}
\begin{equation}
[L_m^{P+Q^{ii}}, L_0^{P+Q^{jj}}] = \frac{1}{f} mL_{m}^{P} +
[L_m^{Q^{ii}}, L_0^{Q^{jj}}] ,
 \label{CRLPQLPQ0}  
\end{equation}
\begin{eqnarray}
&& [L_{m\ne 0}^{Q^{ij}}, L_0^{Q^{kk}}] 
= \frac{1}{2} \frac{\delta_{jk}f-1}{\delta_{ik}f-1 } \,[L_{m}^{Q^{ii}}, L_0^{Q^{kk}}] 
+\frac{1}{2} \frac{\delta_{ik}f-1}{\delta_{jk}f-1 }\, [L_{m}^{Q^{jj}}, L_0^{Q^{kk}}] .
 \label{CRLmijL0} 
\end{eqnarray}
In particular, since $[L_m^{Q^{ii}}, L_0^{Q^{ii}}] =(1-\frac{1}{f}) m L_{m}^{Q^{ii}}$,
\begin{equation}
[L_m^{P+Q^{ii}}, L_0^{P+Q^{ii}}] = \frac{1}{f} mL_{m}^{P} +
\left( 1-\frac{1}{f} \right) m L_{m}^{Q^{ii}}.
\end{equation}

Further, if we define the operators 
\begin{equation}
L_n^{E,m} =\;:\alpha_{n-m}\cdot\alpha_{m}: ,\qquad
L_n^{F^{(ab)},r} =\; :\alpha^{a}_{n-r}\cdot\alpha^{b}_{r}:, \qquad
L_r^{G^{(a)},s } = \;:\alpha^{a}_{s}\cdot\alpha_{r-s}: ,
\label{opsLnmetc}
\end{equation}
we can form a closed algebra (with central extension) by using them as generators by noting that
\begin{equation}
L_n^{E,m} = L_n^{E,n-m} ,\qquad L_n^{F^{(ab)},r}  = L_n^{F^{(ba)},n-r} 
\end{equation}
and
\begin{equation}
L_n^{E} = \frac{1}{2} \sum_{m\in {\bf Z}} L_n^{E,m}  ,\qquad
L_n^{F^{(ab)}} = \frac{1}{2} \sum_{r \in {\bf Z}+\frac{1}{2}}  L_n^{F^{(ab)},r} , \qquad
L_r^{G^{(a)} }  =   \frac{1}{2} \sum_{s\in {\bf Z}+\frac{1}{2}}  L_r^{G^{(a)},s } .
\end{equation}
For example, the right-hand sides of eqs.(\ref{relFF})-(\ref{relFG}) can be written by using the operators given in (\ref{opsLnmetc}).
Also, the following relations are useful for analyzing the physical state conditions:
\begin{equation}
 [L_r^{R^j}, L_0^{Q^{ii}}] =\frac{1}{\sqrt{f}} \left( \delta_{ij} -\frac{1}{f}\right) 
 \sum_{a=2}^f (v_a)_i \sum_{s\in {\bf Z}+\frac{1}{2}} s L_r^{G^{(a)},s},
\label{CRLrQ0}
\end{equation}
\begin{equation}
 [L_r^{G^{(a)},s}, L_0^{Q^{ii}}] =\frac{1}{\sqrt{f}} 
 (v_a)_i\sum_{b=2}^f (v_b)_i  s L_r^{G^{(b)},s}, 
 \label{CRLrsQ0}
\end{equation}
and
\begin{eqnarray}
 [L_{m(\ne 0)}^{Q^{ii}}, L_0^{Q^{jj}}] &=&  \left( \delta_{ij} -\frac{1}{f}\right)
 \sum_{a,b} (v_a)_i (v_b)_j \!\sum_{s\in {\bf Z}+\frac{1}{2}} (m-s) L_m^{F^{(ab)},m-s}
 \\
&=&
 \left( \delta_{ij} -\frac{1}{f}\right) 
 \left(
 L_{m}^{Q^{ij}} 
+ \sum_{s\in {\bf Z}+\frac{1}{2}}
[L_{m-\frac{1}{2}}^{G^{(a)},s},\;L_{\frac{1}{2}}^{G^{(b)},m-s}\,]
\right)
\label{CRQGm0}
\end{eqnarray}
where 
\begin{equation}
[L_r^{G^{(a)},s},\;L_{m-r}^{G^{(b)},t}\,]
= \delta_{m, s+t} (r-s) \,L_{m}^{F^{(ab)}, m-s} +\delta^{a,b} \delta_{s+t,0} \,s \,L_{m}^{E,r-s} .
\end{equation}
%


\section{$\!\!$ General solutions of physical state condition}

\label{app2}
\setcounter{equation}{0}
We give a proof that the space of states satisfying the physical state condition 
given in section~\ref{sec32} is spanned by the states of the form (\ref{generalh1}) and (\ref{generalh2}) with the conditions  (\ref{hcond1})- (\ref{hcond3}).
The condition we have to impose is the set of relations ${\rm (I')}$ for $n=1,2$, ${\rm (I'')}$ for $n=1$, ${\rm (II')}$ (or (\ref{12-2a})) for $r=\frac{1}{2}$ and ${\rm (III)}$ as we discussed in the section~\ref{sec32}. 

We first consider the condition (\ref{12-2a}) for $r=\frac{1}{2}$, that is, $L_{\frac{1}{2}}^{G^{(a)},s} | \phi \rangle=0$.
Since
\begin{equation}
[L_{r}^{G^{(a)},s},\alpha_{-t}^{b,\mu} ] =  s\delta^{a,b}\delta_{s,t}  \alpha_{r-s}^\mu
,\qquad
[L_{r}^{G^{(a)},s}, \alpha_{-t+\frac{1}{2}}^{\mu} ] = (r-s)\delta_{s,r-t+\frac{1}{2}}  \alpha_{s}^{a, \mu},
\end{equation}
the operator $L_{\frac{1}{2}}^{G^{(a)},s}$ gives a non-trivial effect only on a state including the oscillator $\alpha_{-s}^{a,\mu}$ for $s>0$, and $\alpha_{s-1/2}^{\mu}$ for $s<0$. 
In fact, $L_{\frac{1}{2}}^{G^{(a)},s}$ changes the oscillator as
 $\alpha_{-s}^{a,\mu} \rightarrow s \alpha_{-s+\frac{1}{2}}^{\mu} $ for $s>0$, and  $\alpha_{s-\frac{1}{2}}^{\mu} \rightarrow (\frac{1}{2}-s)\alpha_{s}^{a,\mu} $ for $s<0$. 
From this property, possible combination of oscillators $\alpha_{-s}^{a,\mu}$ and $\alpha_{-s+\frac{1}{2}}^{\mu} $ (for $s>0$) or $\alpha_{s-\frac{1}{2}}^{\mu}$ and $\alpha_{s}^{a,\mu} $ (for $s<0$) within a state $| \phi \rangle$ satisfying the condition $L_{\frac{1}{2}}^{G^{(a)},s} | \phi \rangle=0$ can be determined as follows.
For $s\ge \frac{3}{2}$, if there is any $\alpha_{-s}^{a,\mu_a}$ in the state $| \phi \rangle$, then, in order to satisfy $L_{\frac{1}{2}}^{G^{(a)},s} | \phi \rangle=0$, there also should be $f-1$ different type of oscillators $\alpha_{-s}^{b,\mu_b}$ $(b\ne a)$ and $\alpha_{-s+\frac{1}{2}}^{\mu}$, all of which should form the following anti-symmetric combination:  
\begin{equation}
\alpha_{-s}^{2,[\mu_2} \alpha_{-s}^{3,\mu_3} \cdots \alpha_{-s}^{f,\mu_f}
\alpha_{-s+\frac{1}{2}}^{\mu]}.
\end{equation}
Also no other mode $-s$ oscillator $\alpha_{-s}^{a,\mu'}$ could be added in the state.
Similarly, for $-s\le -\frac{1}{2}$, if there is any $\alpha_{-s-\frac{1}{2}}^{\mu}$ in the state $| \phi \rangle$, then, 
in order to satisfy $L_{\frac{1}{2}}^{G^{(a)},-s} | \phi \rangle=0$, 
there should be $f-1$ oscillators  $\alpha_{-s}^{a,\mu_a}$ ($a=2,\cdots f$) and no other $\alpha_{-s-\frac{1}{2}}^{\mu'}$ within the state, and the $f$ oscillators should form the combination
\begin{equation}
\alpha_{-s-\frac{1}{2}}^{[\mu}
\alpha_{-s}^{2,\mu_2} \alpha_{-s}^{3,\mu_3} \cdots \alpha_{-s}^{f,\mu_f]}.
\end{equation}
Note that in this case any number of extra $\alpha_{-s}^{b,\nu}$'s can be included within the state.
Finally, for $s= \frac{1}{2}$, to satisfy $L_{\frac{1}{2}}^{G^{(a)},-s} | \phi \rangle=0$, 
 $ | \phi \rangle$ can include any number $K$ of oscillators $\alpha_{-\frac{1}{2}}^{a_i,\mu_i}$ ($i=1,\cdots, K$).  
However, if we write the corresponding combination as 
 \begin{equation}
h^{a_1 a_2\cdots a_K}_{\mu_1\mu_2\cdots\mu_K}
 \alpha_{-\frac{1}{2}}^{a_1,\mu_1} \alpha_{-\frac{1}{2}}^{a_2,\mu_2}\cdots
  \alpha_{-\frac{1}{2}}^{a_K,\mu_K},
 \end{equation}
the coefficient should satisfy  
$p^{\mu_1}h^{a_1 a_2\cdots a_K}_{\mu_1\mu_2\cdots\mu_K}=0$.
From the above discussion, we see that any state satisfying the condition ${\rm (II')}$  for $r=\frac{1}{2}$, i.e., $L_{\frac{1}{2}}^{G^{(a)},-s} | \phi \rangle=0$ for all $s \in {\bf Z}+\frac{1}{2}$, must have the form  (\ref{generalh1}) and (\ref{generalh2}) with the condition $p^{\sigma_i}
h^{a_1a_2\cdots a_K}_{[\mu_1^M \mu_2^M \cdots  \mu^1_1 \mu^1_2\cdots \mu^1_f] \sigma_1\sigma_2\cdots \sigma_K} =0$.

Next, we impose the condition ${\rm (I'')}$ for $n=1$ on a state $|\phi_{{\rm (II')}_{\!\frac{1}{2}}} \rangle$ satisfying ${\rm (II')}$  for $r=\frac{1}{2}$.
The condition ${\rm (II')}$ can be rewritten as the following simpler form 
\begin{equation}
L_n^{F^{(ab)}} |\phi \rangle = 0 \qquad (a,b=2,\cdots f)
\end{equation}
where $L_n^{F^{ab}}$ is given by (\ref{defLnFab}), and the condition for $r=\frac{1}{2}$ is explicitly given by
\begin{eqnarray}
L_{\frac{1}{2}}^{F^{(ab)}} |\phi_{{\rm (II')}_{\!\frac{1}{2}}} \rangle
 &=& \left(
\frac{1}{2}\alpha_{\frac{1}{2}}^a \cdot \alpha_{\frac{1}{2}}^b
+  \alpha_{-\frac{1}{2}}^{(a} \cdot \alpha_{\frac{3}{2}}^{b)}
+ \alpha_{-\frac{3}{2}}^{(a} \cdot \alpha_{\frac{5}{2}}^{b)}
+\cdots
\right)
|\phi_{{\rm (II')}_{\!\frac{1}{2}}} \rangle
\\
&=&\frac{1}{2}\alpha_{\frac{1}{2}}^a \cdot \alpha_{\frac{1}{2}}^b |\phi_{{\rm (II')}_{\!\frac{1}{2}}} \rangle
=0.
\end{eqnarray}
Here the second equality comes from the  properties of $|\phi_{{\rm (II')}_{\!\frac{1}{2}}} \rangle$.
This gives the traceless condition on each pair of indices corresponding to the coefficients of $\alpha_{-\frac{1}{2}}^{a\mu}$'s as (\ref{hcond2}).

Finally, we consider the condition ${\rm (I')}$ for $n=1,2$ on $|\phi_{{\rm (II')}_{\!\frac{1}{2}}} \rangle$.
From the condition ${\rm (II')}$, the integer mode oscillators of a state $|\phi_{{\rm (II')}_{\!\frac{1}{2}}} \rangle$ has to have a form 
\begin{equation}
|\phi_{{\rm (II')}_{\!\frac{1}{2}}} \rangle= \alpha_{-1}^{[\mu_1^1} \alpha_{-2}^{\mu_1^2} \cdots
\alpha_{-M}^{\mu_1^M]}
\; \left(\mbox{\scriptsize half-integer mode oscillators} \right)
|0,p\rangle.
\end{equation}
Since $L_n^P = \frac{1}{f}L_n^E$ and 
\begin{equation}
[L_n^E, \alpha_{-m}^\nu] = m \alpha_{n-m}^\nu,
\end{equation}
the relations 
\begin{eqnarray}
L_1 |\phi_{{\rm (II')}_{\!\frac{1}{2}}} \rangle
&=&
\alpha_{0}^{[\mu_1^1} \alpha_{-2}^{\mu_1^2} \cdots \alpha_{-M}^{\mu_1^M]}
\; \left(\mbox{\scriptsize half-integer mode oscillators} \right)
|0,p\rangle, 
\\
L_2 |\phi_{{\rm (II')}_{\!\frac{1}{2}}} \rangle
&=&   2 \alpha_{-1}^{[\mu_1^1} \alpha_{0}^{\mu_1^2} \alpha_{-3}^{\mu_1^3}\cdots \alpha_{-M}^{\mu_1^M]} \; \left(\mbox{\scriptsize half-integer mode oscillators} \right)
|0,p\rangle
\end{eqnarray}
hold.
This gives the condition on the coefficient as
\begin{equation}
p^{\mu_1^1}
h^{a_1a_2\cdots a_K}_{[\mu_1^M \mu_2^M \cdots  \mu^1_1 \mu^1_2\cdots \mu^1_f] \sigma_1\sigma_2\cdots \sigma_K}=0
\end{equation}
(or the equivalent condition $ p^{\mu_1^2}
h^{a_1a_2\cdots a_K}_{[\mu_1^M \mu_2^M \cdots  \mu^1_1 \mu^1_2\cdots \mu^1_f] \sigma_1\sigma_2\cdots \sigma_K}=0$.)
This concludes the proof. 

\section{Remark on the light-cone gauge}\label{lightcone}
Here we give a remark on the light-cone gauge. Let us first remind how the light-cone gauge was taken for the ordinary string. After the orthonormal gauge is taken, we utilize residual gauge degrees of freedom which preserve the gauge condition to make light-cone oscillators $\alpha_n^+=0$. Then the Virasoro condition $L_n=0$ together with $\alpha_n^+=0$ become second-class constraint which we can solve it for the remaining light-cone oscillators $\alpha_n^-$ in terms of the transverse modes (or eliminate $\alpha_n^{\pm}$ with Dirac bracket.) Thus only transverse variables are left with us.

Now let us turn to our string junction case. We can easily find reparametrization transformation~(\ref{reparametrization}) which preserves the orthonormal gauge condition~(\ref{orthonormal}). They should satisfy
\begin{equation}
\partial_+\epsilon^{(i)-}=0,\quad \partial_-\epsilon^{(i)+}=0,
\end{equation}
where $\pm$ stands for the worldsheet light-cone directions.
The boundary conditions~(\ref{repara_bc1}) and (\ref{repara_bc2}) restrict them further
\begin{equation}
\epsilon^{(i)\pm}(\tau,\sigma)=\epsilon(\tau\pm\sigma),
\end{equation}
where $\epsilon(\tau)$ is an $i$-independent function which satisfies $\epsilon(\tau)=\epsilon(\tau+2\pi)$.
Then we can use this degree of freedom to make $\alpha_n^+=0$ as in the string case. So we left with $2f-1$ sets of oscillators $\alpha_n^-, \alpha_r^{a+}, \alpha_r^{a-}$ other than transverse modes. Our remaining constraints are $V_n, R_r^a, L_n^{Q^{a-1\,a-1}-Q^{a\,a}}$ whose number of degrees of freedom is also $2f-1$ in total.
These are second-class constraints together with $\alpha_n^+=0$. Thus if we were able to solve them in terms of transverse oscillators, then in principle we could reach the light-cone gauge. But these relations are so complicated that we have not succeeded to explicitly solve them yet. Indeed, even in classical level, the authors in \cite{Plyushchay:1984ja} concluded that it is not possible, though their treatment of remaining second-class constraints is not clear.

Thus, although the number of physical degrees of freedom is likely to coincide with that of light-cone gauge, the structure of physical state may not be the same as the naive truncation to the transverse modes.


\end{document}